\documentclass[osajnl,twocolumn,showpacs,superscriptaddress,10pt]{revtex4-1} 
\usepackage{amsmath,amssymb,graphicx}
\providecommand{\abs}[1]{\left|#1\right|}
\begin{document}

\title{Wave-optics description of self-healing mechanism in Bessel beams}

\author{Andrea Aiello}\email{Corresponding author: andrea.aiello@mpl.mpg.de}
\affiliation{Max Planck Institute for the Science of Light, G$\ddot{u}$nther-Scharowsky-Strasse 1/Bau24, 91058 Erlangen,
Germany}
\affiliation{Institute for Optics, Information and Photonics, University of Erlangen-Nuernberg, Staudtstrasse 7/B2, 91058 Erlangen, Germany}

\author{Girish S. Agarwal}
\affiliation{Department of Physics, Oklahoma State University, Stillwater, Oklahoma 74078, USA}

\begin{abstract}
Bessel beams' great importance in optics lies in that  these  propagate without spreading
and can reconstruct themselves behind an obstruction placed across their path. However, a rigorous wave-optics explanation of the latter property is missing.  In this work we study the reconstruction mechanism by means of a wave-optics description. We obtain expressions for the minimum distance beyond the obstruction at which the beam reconstructs itself, which are in close agreement with the traditional one determined from geometrical optics. Our results show that the physics underlying the self-healing mechanism can be entirely explained in terms of the propagation of  plane waves with radial wave vectors lying on a ring.
%
\end{abstract}

\ocis{(070.7345)   Wave propagation; (050.1940)   Diffraction; (070.3185)   Invariant optical fields; (070.2580)   Paraxial wave optics.}

\maketitle 


In this Letter we present a simple explanation for the self-healing mechanism by which Bessel beams, when partially obstructed, recover their original intensity profile after some distance $z_\text{min}$ from the obstruction.
Bessel beams were theoretically predicted and experimentally demonstrated in the late eighties of the last century by Durnin and coworkers \cite{DurninA,DurninB}. The two salient traits of Bessel beams are  the 
capability of propagating without changing the intensity profile (diffraction-free nature), and  the  remarkable capacity of reconstruct themselves after encountering an obstacle (self-healing mechanism). These characteristics attracted considerable interest in the last three decades and have been the subject of numerous investigations \cite{McGloin05,Jaregui05}.  
In particular, the self-healing property proved to be very useful in research applications such as optical manipulation 
\cite{Arlt01,Gar02}, microscopy \cite{Fahrbach10,Fahrbach12} and quantum communication \cite{McLaren14}. Therefore, the self-reconstruction mechanism has been thoroughly studied mainly by means of numerical simulations
\cite{Bouchal98,Morales07,Vyas11,Rop12}. Only recently an analytical investigation, based on Gaussian optics, has been presented \cite{Chu12}. However, in \cite{Chu12} an explicit expression for the  minimum  reconstruction distance $z_\text{min}$ could not be obtained. It is rather unsatisfactory that the value of this parameter, key to the theory of self-healing mechanism, could be hitherto determined on the ground of geometric arguments only.

In this Letter, we remove this deficiency from the theory by presenting a fully wave-optics characterization of the self-reconstruction process. This approach allows for an evaluation of $z_\text{min}$ grounded on physical, as opposed to geometrical, arguments. 
Moreover, using the Babinet principle \cite{Jackson}, we show that the physics of the self-healing mechanism is simply that of propagation of a plane wave through an aperture.

%
%
To begin with, we recall that  a Bessel beam can be thought as the coherent superposition of plane waves of the form \cite{McGloin05}:
\begin{align}\label{s10}
\psi_\text{pw}(\vec{k}_0\cdot\vec{x}) = A(\vec{k}_0) \exp \bigl( i \vec{k}_0\cdot \vec{x} \bigr) ,
\end{align}
where the amplitude $A(\vec{k}_0) = A \exp \left( i  m \varphi\right)$,  with $m \in \{ 0,\pm 1, \ldots\} $ and the wave vector $\vec{k}_0 = k ( \hat{x} \sin \vartheta_0 \cos \varphi  + \, \hat{y} \sin \vartheta_0 \sin \varphi + \hat{z} \cos \vartheta_0 ) \equiv \vec{k}_{0r} + \hat{z} \, k \cos \vartheta_0$
%
%
%
%
%
are functions of the azimuthal angle $\varphi$ solely, being the modulus $\bigl|\vec{k}_0 \bigr| = k$ and the angular aperture $\vartheta_0$ kept constants. Writing the position vector 
 $\vec{x} = \hat{x} x + \, \hat{y} y  +  \hat{z} z= \vec{r}  +  \hat{z} z$ in cylindrical coordinates $(r, \phi,z)$ as $\vec{x} = \hat{x} r \cos \phi + \, \hat{y} r \sin \phi +  \hat{z} z$, with $r = \sqrt{x^2 + y^2}$, one can write
\begin{align}\label{s30}
\psi_\text{B}(r,  \phi,z) = & \; \int_0^{2 \pi} \psi_\text{pw}(\vec{k}_0 \cdot \vec{x}) d \varphi \nonumber \\
%
%
%
\equiv & \; 2 \pi A \, i^m  \,  e^{ i z \, k_{0z}  } e^{ i  m \phi} J_m \bigl( r  k_{0r} \bigr),
\end{align}
where $k_{0z} = k \cos \vartheta_0$, $k_{0r} = k \sin \vartheta_0$ and $J_m(x)$ denotes the Bessel function of the first kind of order $m$. A straightforward consequence of Eq. \eqref{s30} is that the Fourier transform $\widetilde{\psi}_\text{B}(k_r,  \varphi,0)$ of the Bessel field $\psi_\text{B}(r,  \phi,0)$ is localized on a \emph{ring} of equation $k_x^2 + k_y^2 = k^2 \sin^2 \vartheta_0^2$, namely $\widetilde{\psi}_\text{B}(k_r,  \varphi,0) = 2 \pi A  \exp\left( i  m \varphi \right) \delta \left( \vartheta - \vartheta_0 \right)/{k_{r}^2}$, where $k_r = k \sin \vartheta = (k_x^2 + k_y^2)^{1/2}$. This instance is very different from, e.g., the Fourier transform of a Laguerre-Gauss (LG) beam  whose angular spectrum is essentially localized on the \emph{disc} of equation $k_x^2 + k_y^2 \lesssim k^2  \theta_0^2$, where $\theta_0$ denotes the angular spread of the LG beam.
As we shall see later, this simple fact lies at the foundation of the self-healing mechanism.

%
%
%
%
%
%


Consider a circular opaque object (obstruction) of radius $a$ placed in the $xy$-plane at $z=0$ and characterized by the transmission function
\begin{align}\label{s40}
\tau(r) = & \; \begin{cases}
	1, \quad r > a,\\
	0, \quad r \leq a,
\end{cases} \nonumber \\
= & \; 1 - \Theta \left( a - r\right),
\end{align}
where $\Theta(x)$ denotes the Heaviside step function and $\Theta \left( a - r\right)$, according to the Babinet principle, coincides with the transmission function of an aperture of radius $a$ complementary to the obstacle \cite{Bouchal98}. The Bessel beam at $z=0$ behind the obstacle can be therefore written as
\begin{align}\label{s50}
\psi_\text{B}^O(r,  \phi,  0) = & \; \psi_\text{B}(r,  \phi,  0) \tau(r) \nonumber \\
 = & \; \psi_\text{B}(r,  \phi,  0) - \psi_\text{B}(r,  \phi,  0) \Theta \left( a - r\right)\nonumber \\
 \equiv & \; \psi_\text{B}(r,  \phi,  0) - \psi_\text{B}^A(r,  \phi,  0),
\end{align}
where the superscripts ``$O$'' and ``$A$'' stand for ``\emph{O}bstruction'' and ``\emph{A}perture'', respectively. 
According to a simple ray-tracing model, different authors found  for the minimum reconstruction distance the following expression \cite{McGloin05}:
\begin{align}\label{s60}
z_\text{min} = \frac{a}{\tan \vartheta_0} \equiv z_0.
\end{align}
This means that along and close to the $z$-axis ($r\approx 0$), at any distance $z > z_{{\text{min}}}$ from the obstruction, one should approximately have $\psi_\text{B}^O(r,  \phi,  z) \approx \psi_\text{B}(r,  \phi,  z)$,
%
%
%
%
where $\psi_\text{B}(r,  \phi,  z)$ denotes the Bessel beam that would propagate to $z$ if the obstacle were not present.  Thus, one can estimate $z_{{\text{min}}}$ by evaluating the minimum propagation distance along the  $z$-axis for which it has $\Delta (r,  \phi,  z) \equiv \psi_\text{B}(r,  \phi,  z) - \psi_\text{B}^O(r,  \phi,  z) \approx 0$. According to Eq. \eqref{s50}, such deviation $\Delta$ can be evaluated as
\begin{align}\label{s80}
\Delta(r,\phi,  z) = & \; \psi_\text{B}(r,  \phi,  z) - \left[
\psi_\text{B}(r,  \phi,  z) -   \psi_\text{B}^A(r,  \phi,  z)  \right] \nonumber \\
= & \;    \psi_\text{B}^A(r,  \phi,  z),
\end{align}
where $ \psi_\text{B}^A(r,  \phi,  z)$ denotes the beam transmitted across a circular \emph{aperture} of radius $a$, complementary to the obstruction. Therefore, the whole problem reduces to the calculation of distance along the $z$-axis   where the amplitude of the field $\psi_\text{B}^A(r,  \phi,  z) $  becomes negligible, namely $\psi_\text{B}^A(0,  \phi,  z_\text{min})\approx 0 $. 
However, Eq. \eqref{s60} and experimental results \cite{Bouchal98,McGloin05}, show that $z_\text{min}$ is determined, ceteris paribus, by the angular aperture $\vartheta_0$ solely. Then, since all the plane waves constituting the angular spectrum of a Bessel beam form the same angle $\vartheta_0$ with respect to the $z$-axis (ring domain in $k$-space), it follows that all these waves yield  the same value for $z_\text{min}$. Therefore, since
\begin{align}\label{s90}
\psi_\text{B}^A(r,  \phi,z) = & \; \int_0^{2 \pi} \psi_\text{pw}^A(\vec{k}_0 \cdot \vec{x}) \, d \varphi,  
\end{align}
where $\left.\psi_\text{pw}^A(\vec{k}_0 \cdot \vec{x})\right|_{z=0} = \psi_\text{pw}(\vec{k}_{0r} \cdot \vec{r}) \Theta(a-r)$, in order to determine $z_\text{min}$ it is sufficient to calculate the wave field transmitted by the aperture when the latter is illuminated by the \emph{single} plane wave $\psi_\text{pw}(\vec{k}_0\cdot\vec{x})$. The same reasoning clearly fails for  beams of other forms, as the LG ones, whose angular spectrum is made of plane waves forming different angles $\vartheta \neq \vartheta_0$ (disk domain in $k$-space) with respect to the $z$-axis. In this case, each plane wave determine a different value for $z_\text{min}$ and the latter can take any value. 
This is our first main result. In the remainder we will determine $z_\text{min}$ for the two cases of a square and a soft-Gaussian aperture.

\begin{figure}[h]
\centerline{\includegraphics[clip=false,width=1.2\columnwidth,trim = 0 130 0 90]{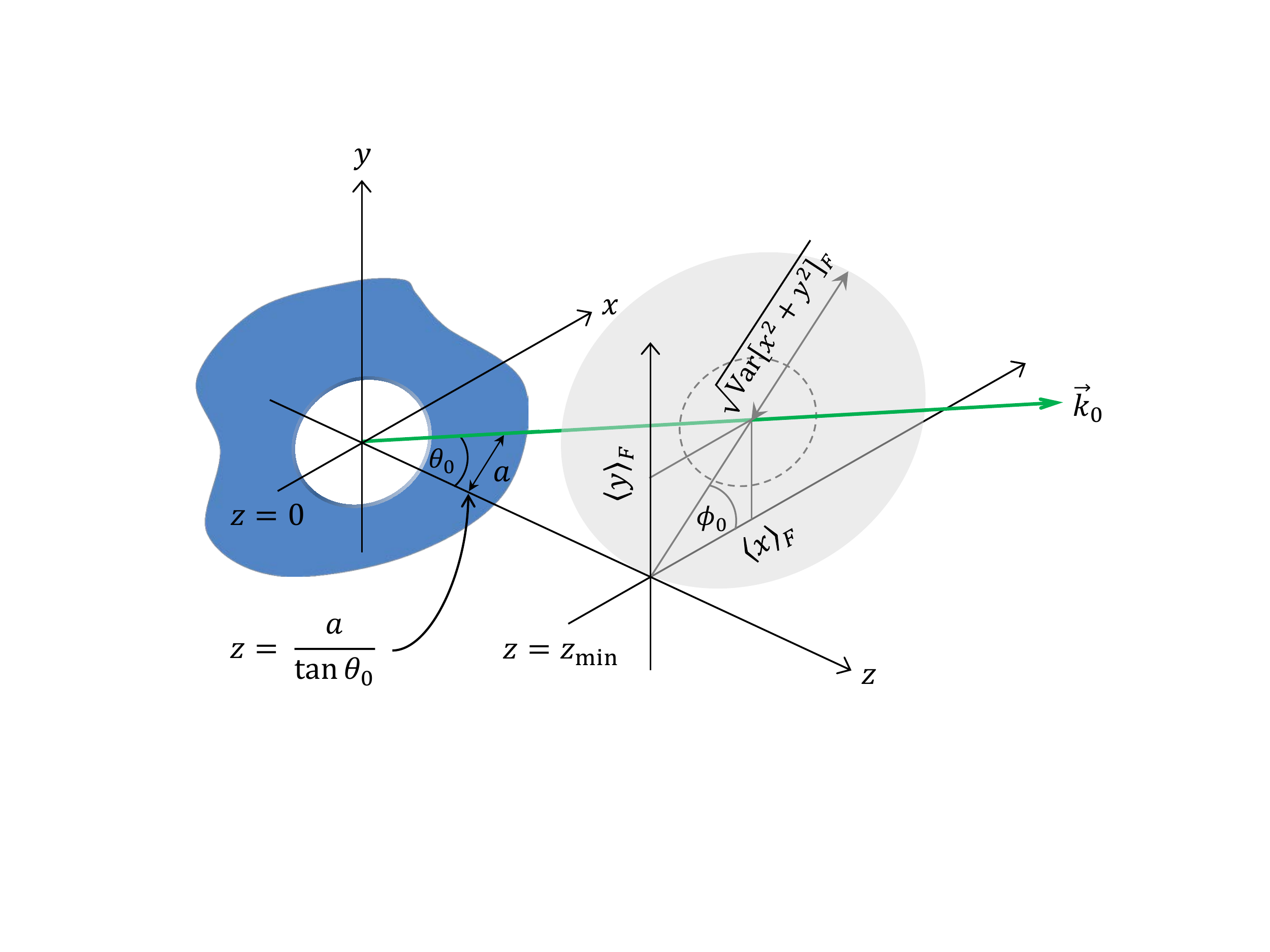}}
\caption{\label{fig1}
The plane wave field $f(x,y) = \exp(i \vec{k}_{0r} \cdot \vec{r})$,  impinges upon an opaque screen with a circular aperture of radius $a$, placed at $z=0$. The  field $f^A(x,y,z)$ transmitted across the aperture  has an intensity distribution $F(x,y,z) \propto \abs{f^A(x,y,z)}^2$ whose spread is represented,  at $z = z_\text{min}$, by the gray area centered at $\langle \vec{r} \rangle_F = \hat{x} \langle x \rangle_F + \hat{y} \langle y \rangle_F$. The extension of this area is quantified by the variance $\operatorname{Var}[r^2]_F = \langle r^2 \rangle_F - \langle \vec{r} \rangle_F \cdot \langle \vec{r} \rangle_F $ of the distribution $F$. The dashed circle represents the geometrical optics image of the aperture upon the plane $z = z_\text{min}$.}
\end{figure}

%
Consider now the scheme illustrated in Fig. \ref{fig1}.
%
%
%
%
%
 Let $f^A(x,y,z) = \psi_\text{pw}^A(\vec{k}_0 \cdot \vec{x})$ be the field transmitted across the  circular aperture of radius $a$. The normalized intensity distribution $F(x,y,z)$ is defined as 
\begin{align}\label{s100}
F(x,y,z)= \frac{\displaystyle{\abs{f^A(x,y,z)}^2}}{\displaystyle{\int\abs{f^A(x,y,z)}^2 \,  d  x  d  y }},
\end{align}
where henceforth integration is always understood upon the whole $xy$-plane if not stated explicitly.
At any plane $z = \text{const.}$, the center of the transmitted wave field can be identified with the  centroid of the intensity  distribution, namely
\begin{align}\label{s110}
\langle \vec{r} \rangle_F = \hat{x} \langle x \rangle_F + \hat{y} \langle y \rangle_F,
\end{align}
where the symbol $\langle \, \cdot \, \rangle_F$ denotes the expectation value with respect to the distribution $F$ 
\begin{align}\label{s120}
\langle h(x,y,z) \rangle_F = \int h(x,y,z) F(x,y,z) \,  d  x  d  y ,
\end{align}
for any function $h(x,y,z)$. At distance $z$ from the screen the diffracted field $f^A(x,y,z)$ spreads over a region whose width can be estimated by the variance of the intensity distribution $F$:
\begin{align}\label{s130}
\operatorname{Var}[r^2]_F  = & \; \langle \vec{r}\cdot \vec{r} \rangle_F - \langle \vec{r} \rangle_F \cdot \langle \vec{r} \rangle_F \nonumber \\
 = & \; \langle x^2 + y^2 \rangle_F -\langle x \rangle_F^2 -\langle y \rangle_F^2.
\end{align}
Both functions $\langle \vec{r} \rangle_F$ and $\operatorname{Var}[r^2]_F $ vary with  $z$. Then, one can (arbitrarily) define $z_\text{min}$ as the distance at which the displacement $\abs{\langle \vec{r} \rangle_F}$ of the centroid of the beam from the $z$-axis equals the half-width of the intensity distribution, that is:
\begin{align}\label{s140}
 \langle \vec{r}\cdot \vec{r} \rangle_F - \langle \vec{r} \rangle_F \cdot \langle \vec{r} \rangle_F =  \langle \vec{r} \rangle_F \cdot \langle \vec{r} \rangle_F, \;\text{at} \; z = z_\text{min} .
\end{align}

To perform this calculation, let us first consider an \emph{arbitrary} field $f(x,y,z)$ that admits a \emph{real-valued} angular spectrum $\widetilde{f}  (k_x,k_y) \in \mathbb{R}$  made of homogeneous plane waves only  \cite{MandelBook}, namely
\begin{align}\label{a10}
f(x,y,z) = \frac{1}{2 \pi} \int \widetilde{f}  (k_x,k_y) e^{ i \left( x k_x + y k_y + z k_z \right) }\,  d^2 k ,
\end{align}
with $k_z = + \left( k^2 - k_x^2 - k_y^2 \right)^{1/2} \geq 0$ and $ d^2 k = d k_x   d  k_y$. Moreover, assume that $f(x,y,z)$ is normalized: $\int\abs{f(x,y,z)}^2 \,  d  x  d  y =1$. Then, it is not difficult to show that it is always possible to write 
\begin{subequations}\label{s160}
\begin{align}
\langle \xi \rangle_F = & \;  z \, \mu_\xi, \label{s160a} \\
\langle \xi^2 \rangle_F = & \; \sigma_\xi^2 + z^2 \, v_\xi^2, \label{s160b}
\end{align}
\end{subequations}
where $\xi \in \{x,y\}$ and 
\begin{subequations}\label{s170}
\begin{align}
\mu_\xi = & \; \int \frac{k_\xi}{k_z} \, \widetilde{f}^{\,2}(k_x,k_y) \,  d  k_x  d  k_y, \label{s170a} \\
\sigma_\xi^2 = & \;  \int  \left[\frac{\partial  \widetilde{f}(k_x,k_y)}{\partial \, k_\xi} \right]^2   d  k_x  d  k_y, \label{s170b} \\
v_\xi^2 = & \; \int \frac{k_\xi^2}{k_z^2} \, \widetilde{f}^{\,2}(k_x,k_y) \,  d  k_x  d  k_y. \label{s170c}
\end{align}
\end{subequations}
Substituting Eqs. \eqref{s160} into Eq. \eqref{s140} yields to a quadratic equation in $z$ whose positive solution is 
\begin{align}\label{s180}
z_\text{min} = \sqrt{\frac{ \displaystyle{\sigma_x^2 + \sigma_y^2}}{ \displaystyle{ 2  \left(\mu_x^2 + \mu_y^2 \right) - \left(v_x^2 + v_y^2 \right)}}} \, .
\end{align}
It should be noticed that this relation is \emph{exact} and does not rely on any approximation. The crucial quantity that uniquely determines $z_\text{min}$ is the ratio 
\begin{align}\label{s185}
\rho(z) = \frac{ \langle \vec{r}\cdot \vec{r} \rangle_F}{  \langle \vec{r} \rangle_F \cdot \langle \vec{r} \rangle_F} = \frac{v_x^2 + v_y^2}{\mu_x^2 + \mu_y^2} + \frac{1}{z^2} \frac{\sigma_x^2 + \sigma_y^2}{\mu_x^2 + \mu_y^2},
\end{align}
with $\rho(z_\text{min}) =2$ by definition. This is our second main result.

In the remainder of this Letter we shall apply Eq. \eqref{s180} to  two relevant cases: \emph{a}) a square aperture of side $2a$ and  \emph{b}) a soft-edge Gaussian aperture with variance $a^2$. Our ultimate goal is to compare the expressions for $z_\text{min}$ obtained from Eq. \eqref{s180} with the geometrical optics one given in Eq. \eqref{s60}.
%
%
%

\begin{figure}[!h]
\centerline{\includegraphics[clip=false,width=.8\columnwidth]{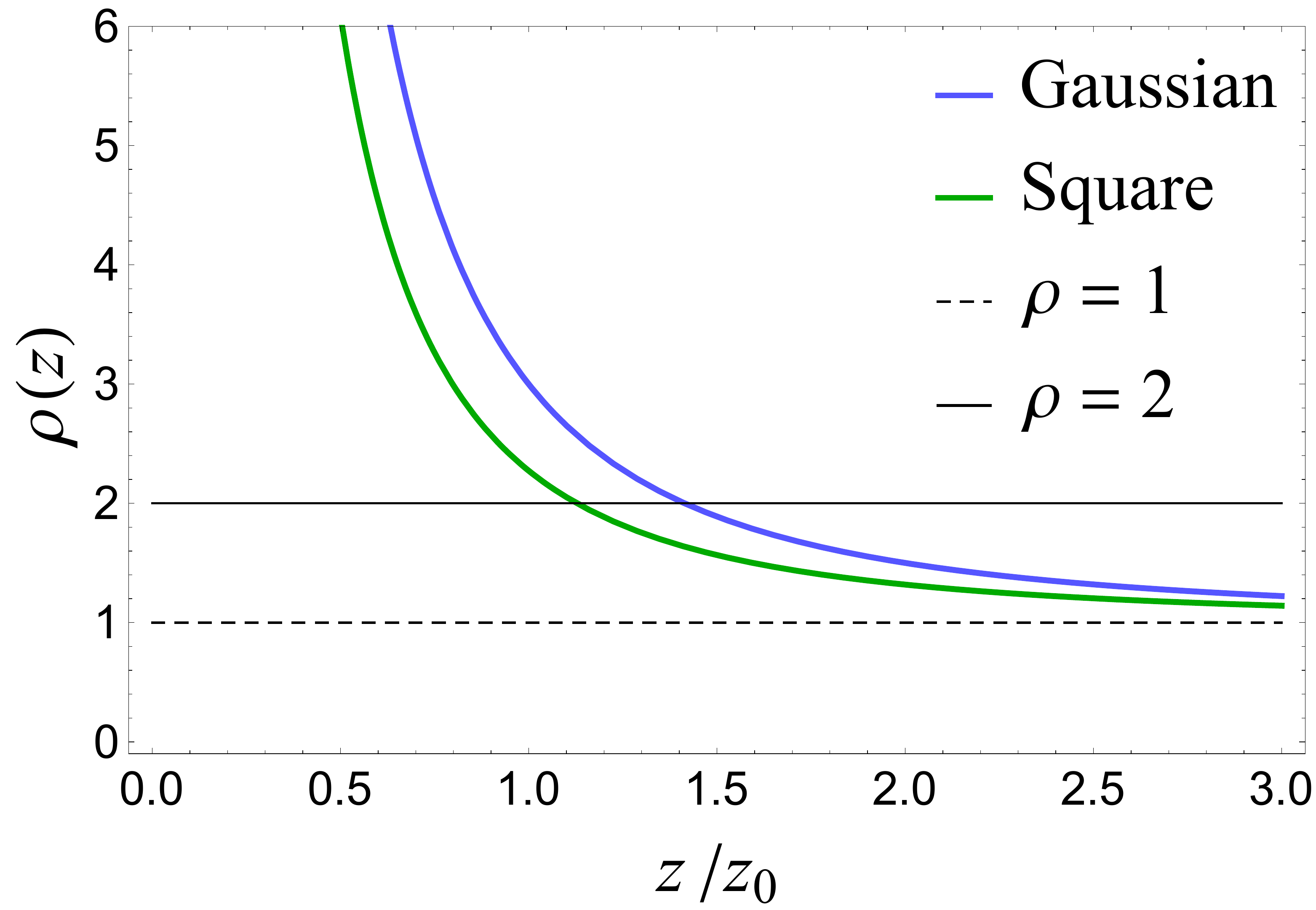}}
\caption{\label{fig2}
Plot of the ratio $\rho(z)$ 
for both the case of square aperture (green line) and of soft Gaussian aperture (blue line), as given in Eqs. \eqref{s265} and \eqref{s320}, respectively. In both case, for $z \to \infty$ this ratio goes to $1$. The parameter $z_0 = a/\tan \vartheta_0$ is the ``geometrical'' value for $z_\text{min}$, as given by Eq. \eqref{s60}.}
\end{figure}

%
\emph{a})  \emph{Square aperture}.  Let us write write explicitly
\begin{align}\label{s190}
\left. \psi_\text{pw}(\vec{k}_0\cdot\vec{x}) \right|_{z=0}= \exp\left[ i ( x k_{0x} + y k_{0y} ) \right] 
\equiv  f(x,y),
\end{align}
where $k_{0x} = k_{0r} \cos \varphi$ and $k_{0y} = k_{0r} \sin \varphi$.
At $z=0$, the field transmitted across the square aperture is written as
\begin{align}\label{s200}
f^A(x,y) = f(x,y)\Theta(a - \abs{x})\Theta(a - \abs{y}).
\end{align}
The Fourier transform is easily calculated:
\begin{align}\label{s210}
\widetilde{f}(k_x,k_y) = & \; \frac{1}{2 \pi} \int_{-\infty}^{\infty} f^A(x,y) \, e^{- i \left( x k_x + y k_y \right) } \,  d  x  d  y \nonumber \\
= & \; \frac{1}{2 \pi} \int_{-a}^{a} e^{- i x (k_x - k_{0x})}  d  x  \int_{-a}^{a} e^{- i y (k_y - k_{0y})}  d  y \nonumber \\
 = & \; 2 \pi  \frac{\sin \left[ a \left( k_x - k_{0x} \right) \right]}{\pi \left( k_x - k_{0x} \right)} \, \frac{\sin \left[ a \left( k_y - k_{0y} \right) \right]}{\pi \left( k_y - k_{0y} \right)}.
\end{align}
In the limit of infinitely wide aperture $a \to \infty$, one recovers the impinging plane wave because of the Dirac delta function definition 
\begin{align}\label{s220}
\lim_{a \to \infty} \frac{\sin \left[ a \left( k_\xi - k_{0\xi} \right) \right]}{\pi \left( k_\xi - k_{0\xi} \right)} = \delta \left( k_\xi - k_{0\xi} \right),
\end{align}
where $\xi \in \{x,y\}$. Now, the key trick is based on the observation that the same delta function can also be realized via a Gaussian function:
\begin{align}\label{s230}
\lim_{a \to \infty} \frac{a}{\pi} \exp \left[ -\frac{a^2}{\pi}\left( k_{0\xi} - k_\xi \right)^2  \right]= \delta \left( k_{0\xi} - k_\xi \right) .
\end{align}
Therefore, for sufficiently large $a \gg 2 \pi/k$ one can approximate $\widetilde{f}(k_x,k_y) $ with 
\begin{align}\label{s240}
\widetilde{f}(k_x,k_y) \approx  2 \pi \frac{ a}{\pi} e^{ -a^2\left( k_{0x} - k_x \right)^2/\pi } \frac{ a}{\pi}  e^{-a^2 \left( k_{0y} - k_y \right)^2/\pi } .
\end{align}
The function in Eq. \eqref{s240} is real, therefore we can apply Eqs. \eqref{s170} and obtain, after a straightforward calculation:
\begin{subequations}\label{s250}
\begin{align}
\mu_x = & \;  \frac{k_{0x}}{k_{0z}} = \tan \vartheta_0 \cos \varphi , \label{s250a} \\
\mu_y = & \;  \frac{k_{0y}}{k_{0z}} = \tan \vartheta_0 \sin \varphi , \label{s250b} \\
\sigma_x^2 = & \; \sigma_y^2 = \frac{2 a^2}{\pi}, \label{s250c} 
\end{align}
\end{subequations}
and $v_\xi^2 = \mu_\xi^2$ where, as usual, $\xi \in \{ x,y \}$.
Substituting these results into Eq. \eqref{s180} leads to:
\begin{align}\label{s260}
z_\text{min} =  ({2}/{\sqrt{\pi}}) z_0,
\end{align}
with $2/\sqrt{\pi} \approx 1.13$. Moreover, from Eq. \eqref{s185} we obtain
\begin{align}\label{s265}
\rho(z) =   1 + \frac{4}{\pi} \left(\frac{z_0}{z} \right)^2 = 1 +  \left(\frac{z_\text{min}}{z} \right)^2.
\end{align}
The plot of this ratio is shown in Fig. \ref{fig2}.
%
%
%

In the evaluation of the integrals in Eqs. \eqref{s170} we have made a Taylor expansion of $k_z$ around $k_z = k_{0z} = k \cos \vartheta_0$ up to and including  first-order terms. Explicitly, after defining the (supposedly small) difference between the transverse parts of the central wave vector $\vec{k}_0$ and the diffracted one $\vec{k}$ as $\vec{q} = \hat{x} (k_{0x} - k_{x}) + \hat{y} (k_{0y} - k_{y})$, one has
\begin{align}\label{s270}
\frac{1}{k_z^{n}} \approx & \; \frac{1}{k_{0z}^n} -n \frac{\vec{k}_0 \cdot \vec{q}}{k_{0z}^{n+2}} + O\left( \frac{q}{k}\right)^2,
\end{align}
with $n=1,2, \ldots$
%
%
%

%
\emph{b}) \emph{Soft-edge Gaussian aperture}. In this case we assume $\tau(r) = \exp \left[ - {r^2}/\left({2 a^2}\right) \right]$ 
and the Fourier transform of $\psi_\text{pw}(\vec{k}_{0r}\cdot\vec{r}) \tau(r)$ reads as
\begin{align}\label{s290}
\widetilde{f}(k_x,k_y) = a^2   e^{-a^2\left( k_{0x} - k_x \right)^2/{2} } \, e^{ -a^2\left( k_{0y} - k_y \right)^2/{2}  }.
\end{align}
This function is again real and we can apply  Eqs. \eqref{s170} to obtain $v_\xi^2 = \mu_\xi^2$, where  $\mu_x^2 + \mu_y^2 = \tan^2 \vartheta_0$  and $\sigma_x^2 =  \sigma_y^2 = a^2$.
Substituting these results into Eq. \eqref{s180} and using Eq. \eqref{s270}, yields to 
\begin{align}\label{s310}
z_\text{min} =  \sqrt{2} \, z_0,
\end{align}
with $\sqrt{2} \approx 1.41$. This result is consistent with both Eq. \eqref{s260} and the geometrical optics result Eq. \eqref{s60}. From the results above and Eq. \eqref{s185} it follows that
\begin{align}\label{s320}
\rho(z) = & \; 1 + 2 \left(\frac{z_0}{z} \right)^2 = 1 +  \left(\frac{z_\text{min}}{z} \right)^2 .
\end{align}
The behavior of $\rho(z)$ is illustrated in Fig. \ref{fig2}. 

In conclusion, we have shown here that the self-healing mechanism manifested by partially obstructed Bessel beams, is entirely determined  by the single plane-wave propagation across an aperture complementary to the obstruction. From a careful analysis of the latter phenomenon, we could ascertain the minimum propagation distance  from the obstacle after which the Bessel beam recover its original intensity profile. Our results, obtained within the framework of wave optics, confirm and extend the traditional ones attained by purely geometrical arguments. Moreover, these results for scalar beams can be extended to vector Bessel beams \cite{Ornigotti13}.

GSA thanks Bob Boyd, Luis Sanchez Soto, Gerd Leuchs for discussions. 
GSA  thanks Gerd also for the great hospitality at MPL-Erlangen where this work was done.

%
%

\newpage

\end{document}